\documentclass[pre,two column,notitlepage]{revtex4-1}
\usepackage[utf8]{inputenc}
\usepackage{amsmath}
\usepackage{amsfonts}
\usepackage{graphicx}
\usepackage{subfig}
\usepackage{xcolor}
\usepackage{csquotes}
\usepackage{gensymb}
\newcommand{\thetaNot}{\theta^{(0)}}
\newcommand{\thetaOne}{\theta^{(1)}}
\newcommand{\thetaTwo}{\theta^{(2)}}
\newcommand{\thetab}{\boldsymbol{\theta}}

\begin{document}
	\title{Spin cones in random-field XY models}
	\author{Rajiv G Pereira, Ananya Janardhanan, and Mustansir Barma }
	\affiliation{TIFR Centre for Interdisciplinary Sciences, Tata Institute of Fundamental Research, Gopanpally, Hyderabad 500046, India}
	\begin{abstract}
We determine the arrangement of spins in the ground state of the XY model with quenched, random
fields, on a fully connected graph. Two types of disordered fields are considered, namely randomly
oriented magnetic fields, and randomly oriented crystal fields. Orientations are chosen from a
uniformly isotropic distribution, but disorder fluctuations in each realization of a finite system lead
to a breaking of rotational symmetry. The result is an interesting pattern of spin orientations, found
by solving a system of coupled, nonlinear equations within perturbation theory and also by exact
numerical continuation. All spins lie within a cone for small enough ratio of field to coupling
strength, with an interesting distribution of spin orientations, with peaks at the cone edges. The
orientation of the cone depends strongly on the realization of disorder, but the opening angle does
not. In the case of random magnetic fields, the cone angle widens as the ratio increases till a critical
value at which there is a first order phase transition and the cone disappears. With random crystal
fields, there is no phase transition and the cone angle approaches $180^\circ$ for large values of the ratio.
At finite low temperatures, Monte-Carlo simulations show that the formation of a cone and its subsequent alignment along the equilibrium direction occur on two different time scales.
	\end{abstract}

	\maketitle
\section{Introduction}
Frozen-in or quenched disorder is known to have strong effects on the thermodynamic
properties of statistical systems. In particular, the interplay of frozen-in randomness with
cooperative interactions can have a profound influence on the nature of ordered states of
spin systems, and lead to new types of patterning. Customarily, theoretical studies are
carried out by averaging over different realizations of disorder in the thermodynamic limit
of the number of spins $N \rightarrow \infty$. However, this procedure can mask
interesting macroscopic patterns that emerge from the competition between quenched
randomness and cooperativity in a system with large but finite $N$.
We show this by explicitly determining the exact ground state of an XY model in the
presence of randomly oriented fields. In each realization of disorder, we demonstrate that
there is an interesting fan-like arrangement of spins with a robust cone angle, and an
interesting distribution of spins within the cone.

The question of magnetic ordering in the presence of random fields has a long history, and
remains a question of current interest. In this paper we study XY ferromagnets with long-
range interactions, subject to two distinct types of locally quenched fields: random magnetic
fields which promote order in different directions (the RFXY model), and random crystal
fields which define a random set of easy axes for spins (the RCXY model). The RFXY model has been used to study disordered
superconductors \cite{Larkin_1970, MFisher_1989, Nattermann_1991} and crystal surfaces with quenched disorder~\cite{Toner_1990}. On the other hand, the RCXY model was proposed \cite{Harris_1973} to explain the
magnetic properties of amorphous materials, exemplified by rare earth compounds such as
$DyCu$ and $TbAg$~\cite{Coey_1978, Ferrer_1978}. The randomness of anisotropy axis orientation competes with
ferromagnetic exchange and reduces the magnetization \cite{Callen_1977, footnote1}. The model
has been used to understand the magnetic properties of many amorphous binary alloys \cite{Krey_1977, Alben_1978, Dudka_2005, Idzikowski_2005}, as
well as nanocrystalline \cite{Hernando_1995, Alvarez_2020} and molecular \cite{Girtu_2000} magnets.

The occurrence and nature of order in these systems is a central question. A general
argument was given by Imry and Ma to show that a ferromagnetically ordered state cannot
be sustained in less than two (four) dimensions for discrete (vector) spins \cite{Imry_1975}.
While a spin-glass phase is ruled out in the random-field Ising model \cite{Krzakala_2010},
there is no analogous result which rules out a spin-glass phase in models with continuous
spins. In fact there is evidence
supporting the existence of such a phase in RFXY models with short-range interaction~\cite{Cardy_1982, Doussal_1995}. The RCXY model with short-range interaction is expected to exhibit a spin glass phase below six dimensions~\cite{Aharony_1975, Pelcovits_1978, Chudnovsky_1986}. A spin glass phase is also
found in a Monte Carlo study of the 3D RCXY in infinitely strong crystal fields, in which limit
it reduces to a quenched random-bond Ising model with correlated random couplings of either sign
\cite{Jayaprakash_1980}. However, in higher dimensions, we expect ferromagnetism to prevail at low
temperatures.

The problem for either sort of random field has been studied earlier on fully connected
graphs. Infinite-range connectivity has the advantage that the free energy can be calculated exactly in the
thermodynamic limit. For the RFXY model this was done recently using large deviations
\cite{Sumedha_2022}, belief propagation \cite{Lupo_2019} and replicas \cite{Morone_2023}. The nature of
the critical locus which separates the disordered phase from an ordered phase with finite
magnetization depends on the distribution of random fields: with a Gaussian distribution, it
is second-order throughout \cite{Morone_2023}, while with randomly oriented fields of fixed
magnitude (the case of interest to us), the transition becomes first order for strong fields
and low temperature \cite{Sumedha_2022}. For the RCXY model, there is a similar phase transition to an ordered state, but
the transition locus remains second order throughout \cite{Derrida_1980,Barma_2022}.

Although bulk thermodynamic quantities were calculated in these treatments, the nature of
spin ordering in the ferromagnetic state remained unknown even in the ground state. This is
the central question we address in this paper. With randomly-pointing fields of constant
magnitude drawn from an isotropic distribution, in a typical configuration, an effective field of order $1/{\sqrt N}$
breaks the isotropic symmetry of the ground state. Minimization of the energy yields $N$
coupled nonlinear equations for the spin orientations. We develop a perturbative method to
solve for the spin angles for small values of the ratio of field strength to the exchange, and
find that they lie within a well-defined cone whose orientation varies from sample to
sample, but whose opening angle is robust. Interestingly, the distribution of spins within the
cone is found to be largest near the cone edges. An exact numerical treatment of
the nonlinear equations confirms the perturbative results and further shows that with
increasing field, there is a first order phase transition to a disordered state, as predicted in
\cite{Sumedha_2022}. The cone angle increases continuously until the transition, beyond which
the cone disappears. On the other hand, in the random crystal field problem, the cone angle
increases continuously to 180 degrees as the strength of the crystal field goes to $\infty$.

We also performed Monte Carlo simulations for both RFXY and RCXY models at low
temperatures, and found that cone formation is robust. An interesting dynamics governs
cone formation and settling: Formation of a well-defined cone is quick and happens on a
time scale determined by the exchange coupling, while its orientation relaxes on a slower
time scale set by the strength of the random field. We develop a phenomenological
equation which describes this dynamics.

\section{The model}
The RFXY model is defined by the Hamiltonian	
\begin{align}\label{H_with_h}
	H_{RF} = -\frac{J}{2 N} \left( \sum_{i=1}^{N} \boldsymbol{S}_i \right)^2
	- h \sum_{i=1}^{N} \boldsymbol{n}_i \cdot \boldsymbol{S}_i,
\end{align}
where the spin $\boldsymbol{S}_i$ is a two dimensional unit vector associated with the $i$th site of a fully connected graph with $N$ sites.  The spin at  site $i$ is coupled to that at site $j$ with an energy $-\frac{J}{N} \boldsymbol{S}_i \cdot \boldsymbol{S}_j$.  At every site there is a random field with site-independent strength $h$ and site-dependent direction determined by the unit vector $\boldsymbol{n}_i$. The unit vectors $\{\boldsymbol{n}_i\}$ are independent and identically distributed random vectors that can lie anywhere on a unit circle with equal likelihood. The spin at site $i$ is coupled to the random field at that site with an energy $-h \boldsymbol{n_i}\cdot \boldsymbol{S}_i$.

The RCXY model is obtained by replacing the last term in the Hamiltonian~(\ref{H_with_h}):
\begin{align}\label{H_with_D}
	H_{RC} = -\frac{J}{2 N} \left( \sum_{i=1}^{N} \boldsymbol{S}_i \right)^2
	- D \sum_{i=1}^{N} \left(\boldsymbol{n}_i \cdot \boldsymbol{S}_i \right)^2,
\end{align}
where $D$ represents the strength of the random crystal field in the direction $\boldsymbol{n}_i$ at site $i$. The unit vectors $\{\boldsymbol{n}_i\}$ are distributed uniformly as in the RFXY model. The RCXY model has an underlying Ising symmetry: the Hamiltonian is invariant under the transformation $\left\{\boldsymbol{S}_i \,|\, i=1, \; 2, ...N \right\} \rightarrow  \left\{-\boldsymbol{S}_i \,|\, i=1,\;2,...N\right\}$.

As mentioned earlier, the RFXY model exhibits a locus of order-disorder phase transitions in the $(T/J, h/J)$ plane, where $T$ is the temperature. 
The transitions are of first order in the portion of the locus lying in the low temperature regime, which is separated from the portion lying in high temperature regime, where the transitions are continuous, by a tricritical point.
For the RCXY model the transitions are of second order everywhere and the critical temperature is independent of the ratio $D/J$~\cite{Derrida_1980, Barma_2022}.  Further, there is no phase transition to a disordered state when the ratio of field to coupling strength is varied at $T=0$, in contrast to the RFXY model.

In the following sections we will be interested in the nature of the ordered state in both these models at  $T=0$ and at temperatures close to zero. In particular, we will explore the arrangement of spins in a finite system with a given configuration of fields. Though the field-orientations are chosen from an isotropic distribution, rotational symmetry is broken for any finite $N$.  

\section{RFXY: Cones at zero-temperature}\label{sec:3}	
In this section, we explore the $T=0$ arrangement of spins for the RFXY model.

	\subsection{Perturbation theory: $h/J<<1$}
An insight into the distribution of spins can be obtained by analyzing the ground state of the Hamiltonian~(\ref{H_with_h}) using perturbation theory. To this end, Eq.~(\ref{H_with_h}) is first recast in terms of the angles $\{\theta_i\}$ and $\{\alpha_i\}$ that the spins $\{\boldsymbol{S}_i\}$ and the fields $\{\boldsymbol{n}_i\}$, respectively, make with the $x$-axis.  The ensuing equation is then extremized with respect to $\theta_i$, yielding the following set of $N$ equations for the ground state. 
\begin{align}\label{H_with_h_extrema}
		\frac{1}{N} \sum_{j=1}^{N} \sin(\theta_i - \theta_j)
		+ \frac{h}{J} \sin(\theta_i - \alpha_i)=0.
\end{align}
To solve these, we expand $\theta_i$ in powers of $h/J$ as
\begin{align}\label{expansion_form_1}
	\theta_i = \thetaNot + \frac{h}{J} \, \thetaOne_i + \frac{h^2}{J^2} \, \theta_i^{(2)} + ...
\end{align}
Note that the $0$'th  order term $\thetaNot$ is independent of $i$, since all spins point in same direction in the limit $h/J \rightarrow 0$. On substituting Eq.~(\ref{expansion_form_1}) in Eq.~(\ref{H_with_h_extrema}) and then expanding in powers of $h/J$, we obtain the following equations at $\mathcal{O}(h/J)$ and $\mathcal{O}(h^2/J^2)$, respectively:
\begin{align}
	\sum_{j=1}^{N} \left(\thetaOne_i - \thetaOne_j\right) + N\sin(\thetaNot - \alpha_i) &=0, \label{first_order_h}	
	\\ 
	 \sum_{j=1}^{N} \left(\thetaTwo_i - \thetaTwo_j\right)  + N  \thetaOne_i \cos(\thetaNot-\alpha_i)&=0.
	 \label{second_order_h}
\end{align}
Note there are $N$ equations at each order. On summing the \enquote*{first-order} equations (\ref{first_order_h}) we obtain the $0$'th order contribution to $\theta_i$ as
\begin{align}
	\sum_{i=1}^{N} \sin(\thetaNot - \alpha_i)&=0 \label{constraint} 
	\\ 
	\implies \tan(\thetaNot) &= \frac{\sum_{i=1}^{N} \sin(\alpha_i)}{\sum_{i=1}^{N} \cos(\alpha_i)}.
	\label{thetaNot}
\end{align}
This implies $\thetaNot = \alpha_0$ or $\alpha_0 + \pi$, where $\alpha_0$ is the angle made by 
\begin{equation}
\boldsymbol{n}_0 \equiv \sum_{i=1}^{N} \boldsymbol{n}_i
\end{equation}
 with the $x$-axis. The latter solution has a higher energy and is therefore discarded. We conclude that at $T=0$, $h=0$ all the spins $\boldsymbol{S}_i$ point in the direction of the vector sum of the fields. Note that for any finite system size $N$, the sum of disordered fields $h \sum_{i=1}^{N} \boldsymbol{n}_i$ is of the order $\sqrt{N}$, which results in singling out a preferred direction $\thetaNot = \alpha_0$, breaking rotational symmetry. 

\begin{figure*}
	\centering
	\subfloat[]
	{
		\includegraphics[scale=0.30]{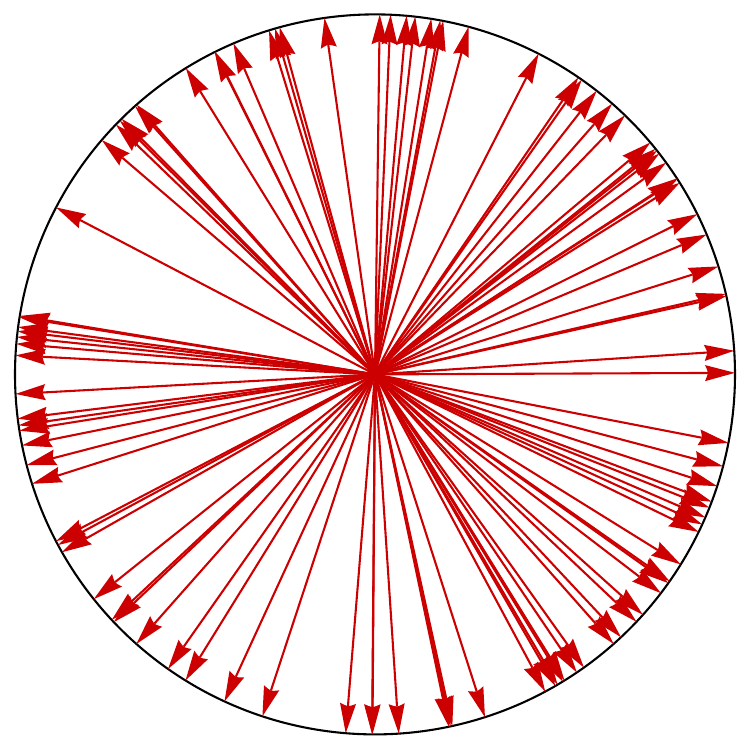}
		\label{subfig:betaArrows_100_3}
	}
	\subfloat[]{
		\includegraphics[scale=0.30]{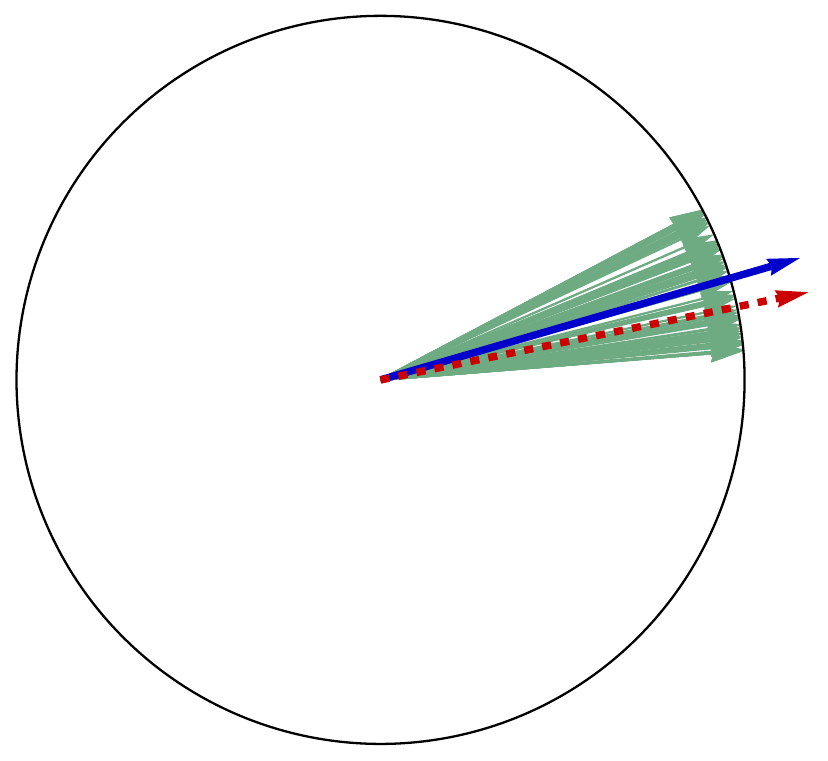}
		\label{subfig:coneAt_0p2_100_3}
	}
	\subfloat[]{
		\includegraphics[scale=0.30]{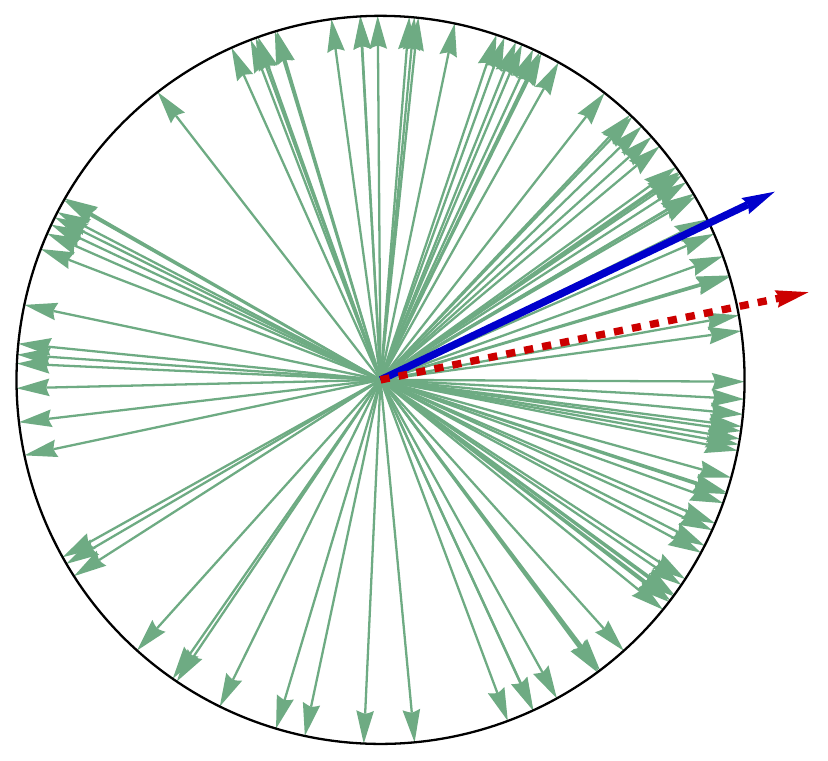}
		\label{subfig:coneAt_0p66_100_3}
	}
	\subfloat[]{
	\includegraphics[scale=0.4]{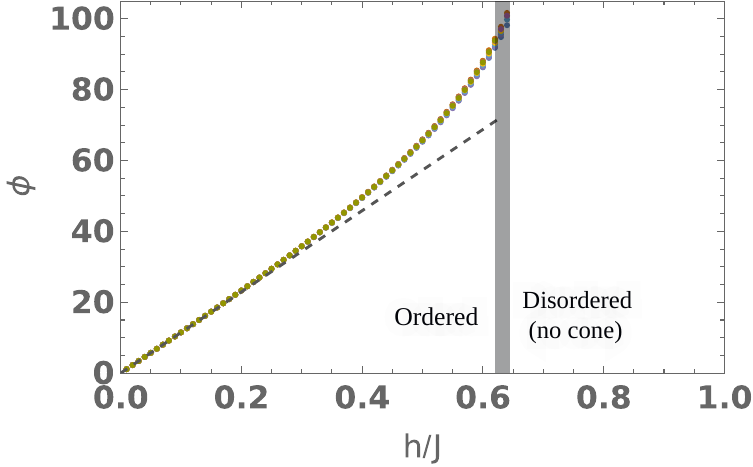}
	\label{fig:phic_vs_h}
}
\\
	\subfloat[]{
		\includegraphics[scale=0.4]{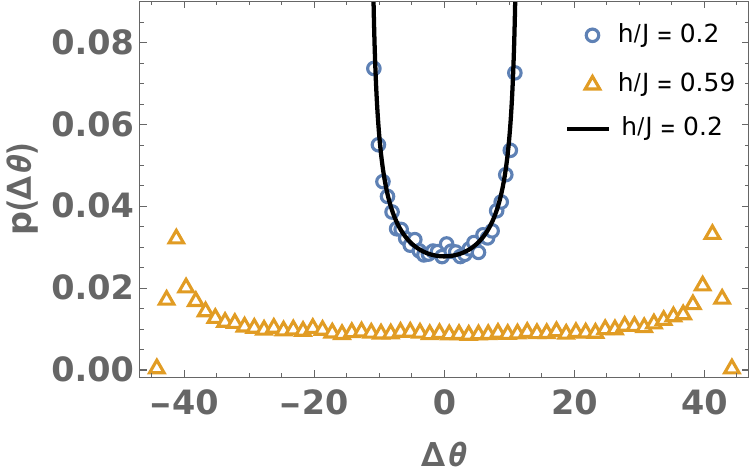}
		\label{fig:hist_h0p2}
	}
	\subfloat[]{
	\includegraphics*[scale=0.4]{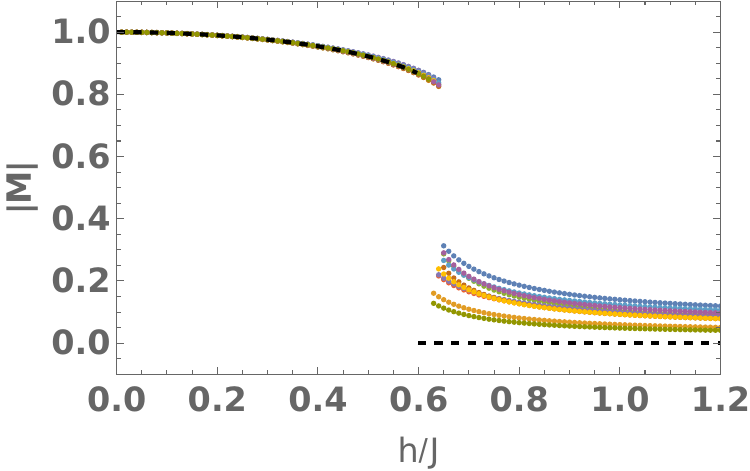}
	\label{fig:mag_vs_h/J}}
	\subfloat[]{
	\includegraphics[scale=0.6]{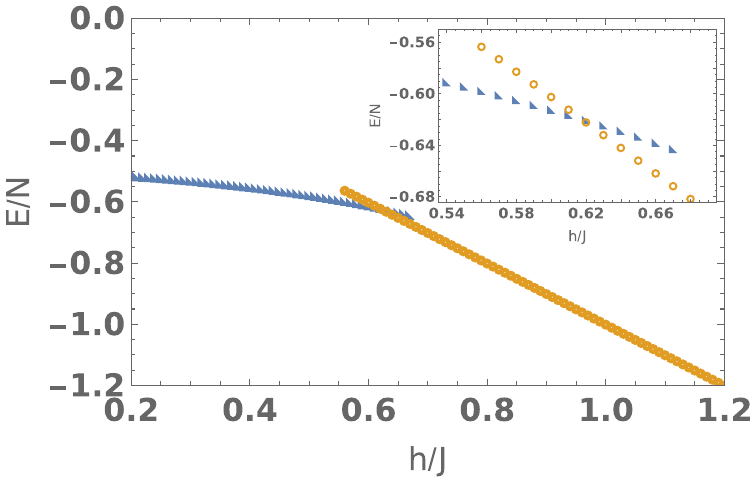}
	\label{fig:meta}}
		\caption{The RFXY model: The sub-figures~(a) and (b) show, respectively, the directions of $\{\boldsymbol{n}_i\}$ and  $\{\boldsymbol{S}_i\}$ for $h/J=0.2$. (c):~$\{\boldsymbol{S}_i\}$ for $h/J=0.63$.  All the three sub-figures are for a typical sample with $N=100$. The protruding red dashed (blue straight) arrow shows the direction of $\boldsymbol{n}_0 = \sum_{i=1}^{N} \boldsymbol{n}_i$  ($\boldsymbol{M}$). (d): Cone angle $\phi$ versus $h/J$ for $10$ different samples of $N=400$. The black dashed line shows cone angle $\phi(h/J)$ according to perturbation theory~(\ref{pert_coneangle}). (e): Distributions of spins at $h/J=0.2$ and $0.59$. The angles are in degrees. The black line shows the probability distribution function~(\ref{probDistn}) for $h/J =0.2$. (f): Magnetization vs $h/J$ for $10$ different samples of $N=400$. The black dashed line shows exact $|\boldsymbol{M}|$ obtained from the large deviation theory~\cite{Sumedha_2022}. (g):
		Energy versus $h/J$ for solutions that minimize the Hamiltonian~(\ref{H_with_h}). Notice the occurrence of metastability.   These results are obtained using the numerical techniques discussed in Sec.~\ref{sec:numc1}. } 
	\label{fig:arrows_h}
\end{figure*}
We now focus on the $N$ first order corrections $\{\thetaOne_i\}$, which satisfy the $N$ equations~(\ref{first_order_h}). However, only $N-1$ of these equations are linearly independent since the equations~(\ref{first_order_h}) add to give $0$. The \enquote*{missing} equation which, together with Eq.~(\ref{first_order_h}), uniquely determines $\thetaOne_i$ is obtained from the \enquote*{second-order} equations~(\ref{second_order_h}), which on summing over $i$ yield $\sum_{i=1}^{N} \thetaOne_i \cos(\thetaNot-\alpha_i) = 0.$ Solving this simultaneously with Eqs.~(\ref{first_order_h}) yields
\begin{align}\label{firstOrderh}
	\thetaOne_i= \sin(\alpha_i - \alpha_0) + \eta
\end{align}
where we have replaced $\thetaNot$ by $\alpha_0$ and 
\begin{align}\label{eta}
	\eta =  \frac{1}{2}
	\frac{\sum_{j=1}^N \sin \left[2 \left( \alpha_0 - \alpha_j\right)\right]}{\sum_{j=1}^N \cos \left( \alpha_0 - \alpha_j\right)}.
\end{align}
Replacing $\thetaOne_i$ in Eq.~(\ref{expansion_form_1}) by Eq.~(\ref{first_order_h}) we get
\begin{equation}\label{soln_perth}
	\theta_i \simeq \alpha_0 + \frac{h}{J} \left(  \sin(\alpha_i - \alpha_0) + \eta \right)
\end{equation}
to first order in $h/J$. 

To find the orientation of the magnetization $\boldsymbol{M}$ and the distribution of spins around it to first order we write 
\begin{align} \label{magh}
	\boldsymbol{M} & \equiv \frac{1}{N} \sum_{i=1}^{N} \boldsymbol{S}_i= \frac{1}{N} \sum_{i=1}^N 
	\left[
	\cos(\theta_i) \boldsymbol{\hat{x}} + \sin (\theta_i) \boldsymbol{\hat{y}}
	\right] \nonumber \\
	& \simeq \boldsymbol{M}^{(0)} - \frac{h}{NJ} \left(
	\sin(\alpha_0) \boldsymbol{\hat{x}} - \cos(\alpha_0)
	\boldsymbol{\hat{y}} \right)\times\sum_{i=1}^{N}  \thetaOne_i,
\end{align}
where $\boldsymbol{M}^{(0)} = \cos(\alpha_0) \boldsymbol{\hat{x}} + \sin(\alpha_0) \boldsymbol{\hat{y}}$ is the magnetization in the limit $h/J \rightarrow 0$. To obtain the last line, we used Eq.~(\ref{soln_perth}) and expanded the resulting expression to first order in $h/J$. The factor in the second term of Eq.~(\ref{magh}) can be evaluated using Eqs.~(\ref{firstOrderh}), (\ref{eta}), and (\ref{constraint}) to obtain $\sum_{i=1}^{N} \thetaOne_i = N \eta$. The orientation $\theta_0$ of the magnetization $\boldsymbol{M}$ is thus
\begin{align}\label{thetaSubNot}
	\theta_0 \simeq \alpha_0 + \frac{h}{J} \eta
\end{align}
to first order in $h/J$. The angle $\Delta \theta_i$ that the $i$th spin makes with $\boldsymbol{M}$ is
\begin{align}
	\Delta \theta_i \equiv \theta_i - \theta_0 \simeq \frac{h}{J} \sin(\alpha_i - \theta_0),
\end{align}
where we have used Eqs.~(\ref{expansion_form_1}), (\ref{firstOrderh}), and (\ref{thetaSubNot}). For random fields that are uniformly distributed over a circle, we obtain the probability distribution for $\Delta \theta_i$ as 
\begin{align}\label{probDistn}
	p(\Delta \theta_i) = \frac{1}{\pi \sqrt{(h/J)^2 - \Delta \theta_i^2}}.
\end{align}
A plot of the distribution $p(\Delta \theta_i)$ for $h/J = 0.2$ is shown by the black line in Fig.~\ref{fig:hist_h0p2}. Note that all the spins lie within a cone whose edges are given by $\Delta \theta_i = \pm h/J$, at which points the distribution diverges. Within the cone, the distribution is minimum at the center. The arrangement of spins in the perturbative regime, for a particular sample of size $N=100$, is depicted using the green arrows confined within the circle in~Fig.~\ref{subfig:coneAt_0p2_100_3}. The red dashed arrow shows the direction of $\boldsymbol{n}_0$ and the blue continuous arrow shows the direction of $\boldsymbol{M}$. The spin configuration shown therein corresponds to the particular field configuration shown in Fig.~\ref{subfig:betaArrows_100_3}. 

 Thus we see that spins are distributed within a cone whose orientation is given by Eq.~(\ref{thetaSubNot}), while the cone angle, defined as the angular separation between the spins at the two farthest edges of the cone, is given by 
\begin{align}\label{pert_coneangle}
	\phi =  \frac{2h}{J},
\end{align}
which is evident from Eq.~(\ref{probDistn}). The cone orientation depends on the configuration of fields, while the cone angle does not.  

\subsection{Numerically continuing to higher $h/J$}\label{sec:numc1}
We now investigate how the conical arrangement of spins changes as $h/J$ increases. We will show that the cone angle increases continuously until a critical value $h_c/J$, at which point there is a first order transition to a disordered state. As $N \rightarrow \infty$ the critical value $h_c/J \rightarrow 0.597$, the value obtained from the large deviation calculation~\cite{Sumedha_2022}.

The $T=0$ distribution of spins is obtained by solving the equation for extremum~(\ref{H_with_h_extrema}) using the method of numerical continuation~\cite{allgower2003introduction}. The equation is independently solved for several configurations of quenched random fields and for different values of $N$. In particular, we choose $N=100, \; 200, \; 300$,  $400$, $600$, and $800$. The procedure is briefly explained below. 

First, $N$ angles $\{\alpha_i\}$ are chosen independently from the interval $\left. (0, 2 \pi \right]$ with uniform likelihood. This fixes the directions of the quenched random fields $\{\boldsymbol{n}_i\}$. Now, the formula derived using the perturbation theory~(\ref{soln_perth}) is used to obtain an approximate solution to Eq~(\ref{H_with_h_extrema}) for a small value of $h/J$ (say, $0.001$). This  solution is used as an initial guess using which we numerically continue to higher values of $h/J$. (See Appendix~\ref{sec:app1} for further details.) 

For small enough $h/J$, we find that the distribution of spins is consistent with the perturbative results. The spins are confined within a cone centered along the direction of magnetization as shown in Fig.~\ref{subfig:coneAt_0p2_100_3}. The number of spins is largest at the edges, where there is a sharp cut-off, and smallest at the center of the cone. This is demonstrated using the case of $h/J = 0.2$ in Fig.~\ref{fig:hist_h0p2} (blue circles), where the probability density $p(\Delta \theta)$ is plotted against $\Delta \theta = \theta -\theta_0$. The probability density $p(\Delta \theta)$ was obtained by sampling over $500$ configurations with $N=100$.

We find that the cone angle $\phi$ widens until the critical value $h_c/J$ is reached.  Figure~\ref{fig:phic_vs_h} shows the variation of $\phi$ for $N=400$.
  The values of the cone angles at the phase transition are found to be $\phi_c = 111 \pm 14, \; 103 \pm 8, \; 101 \pm 6$, $ 97 \pm 4$, $95 \pm 4$, and $94 \pm 4$ degrees for $N=100, \; 200, \; 300,$ $400$, $600$, and $800$, respectively, on averaging over $50$ samples for each $N$.
  The results are consistent with $\phi_c \rightarrow 90^\circ$ as $N \rightarrow \infty$. The distribution of spins within the cone just before the phase transition (at $h/J = 0.59$) is found to be flat around the center of the cone, unlike for small $h/J$ (see Fig.~\ref{fig:hist_h0p2}).  
  The value of $h_c/J$ shows sample to  sample fluctuation and a systematic fall as $N$ increases. We find $h_c/J = 0.662 \pm 0.024, \; 0.645 \pm 0.020, \; 0.638 \pm 0.019$, $0.628 \pm 0.013$, $0.623 \pm 0.013$, and $0.619 \pm 0.013$ for  $N=100, \; 200, \; 300,$  $400$, $600$, and $800$, respectively. Note that the results are consistent with $h_c/J \simeq 0.597$ for $N \rightarrow \infty$.

The numerical continuation scheme fails at a value of $h/J$ slightly greater than $h_c/J$ as there is no ordered state solution to  Eq.~(\ref{H_with_h_extrema}) at this point.  The solutions that minimize the energy~(\ref{H_with_h}) at points beyond $h_c/J$ are the disordered ones.
 To access these, we numerically continue backwards in $h/J$ from the state in the limit $J/h <<1$. The starting guess solution to Eq.~(\ref{H_with_h_extrema}), obtained perturbatively, is $\theta_i \simeq \alpha_i + \frac{J}{N h} \sum_{j=1}^{N} \sin(\alpha_j - \alpha_i)$. 
  
  There is an abrupt change in the arrangement of spins as $h/J$ is increased past the transition point: from being confined to a cone, for $h/J < h_c/J$, they spread over a circle. 
  The distribution of spins in the disordered phase for a typical configuration of fields is shown in Fig.~\ref{subfig:coneAt_0p66_100_3} (at $h/J = 0.65$). We note here that for this sample the phase transition occurs between $h/J =0.64$ and $0.65$.
  
  There is also a sudden jump in the magnetization at the phase transition for each sample, which is clear from the plot in Fig.~\ref{fig:mag_vs_h/J}, indicating clearly that the system is undergoing a first order phase transition. The dashed black line therein shows the exact modulus of magnetization $|\boldsymbol{M}|$ in the thermodynamic limit, calculated using large deviation theory in~\cite{Sumedha_2022}.

 The energies of the minima obtained for a typical configuration of fields using the scheme described above are plotted against $h/J$ in Fig.~\ref{fig:meta}.  The cones and circles used as markers therein indicate that the solutions in these regions correspond to conical and disordered arrangements of spins, respectively. Note that the conical (disordered) states continue to the right (left) of the transition point, where they are no longer the global minima but are still metastable local minima of the Hamiltonian.

\begin{figure*}
	\subfloat[]{
		\includegraphics[scale=0.28]{beta_200_4.pdf}
		\label{subfig:distn_fields}
	}
	\subfloat[]{
		\includegraphics[scale=0.32]{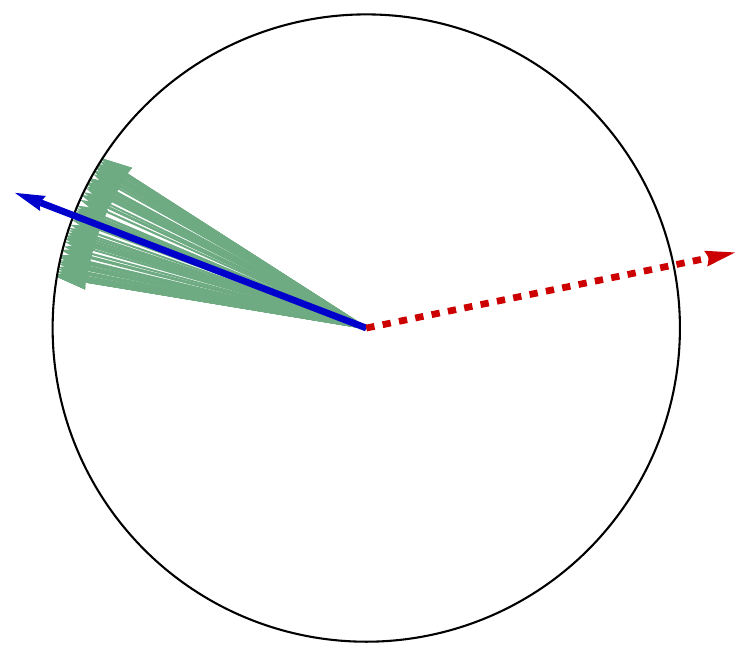}
		\label{subfig:coneAt_d0p2_100_3}
	}
	\subfloat[]{
		\includegraphics[scale=0.32]{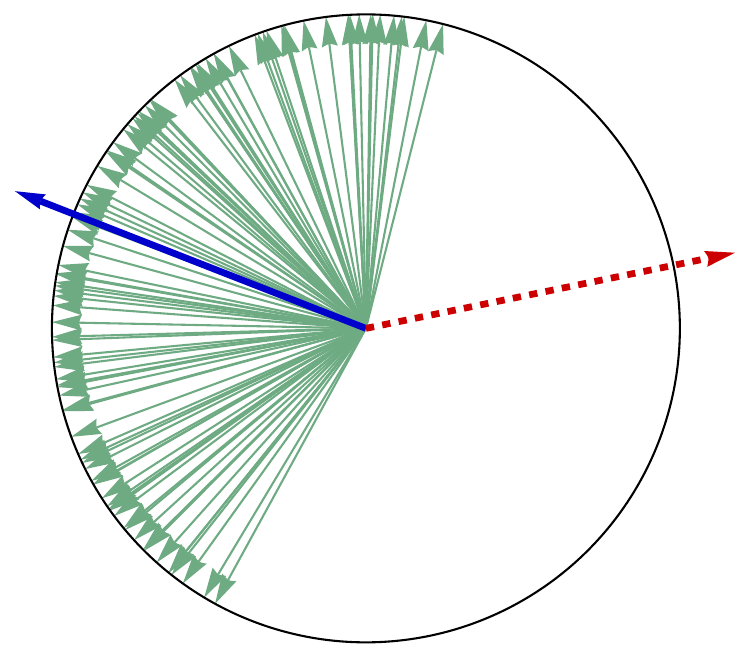}
		\label{subfig:coneAt_d0p66_100_3}
	}
	\subfloat[]{
		\includegraphics[scale=0.4]{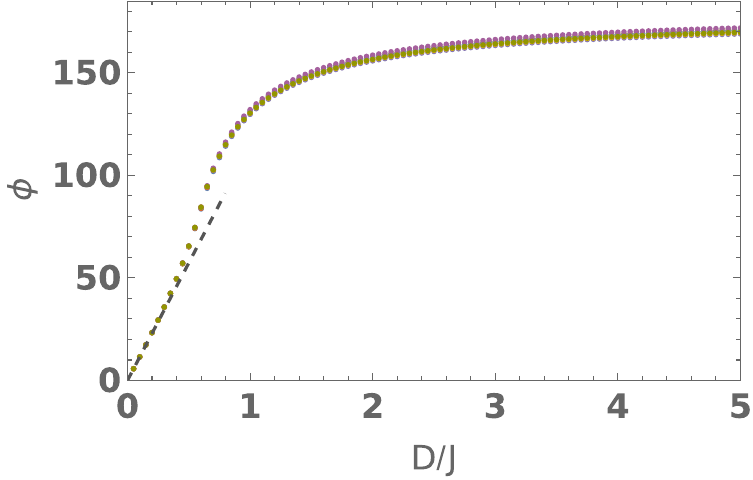}
		\label{fig:phic_vs_d}}
	\\
	\subfloat[]{
		\includegraphics[scale=0.53]{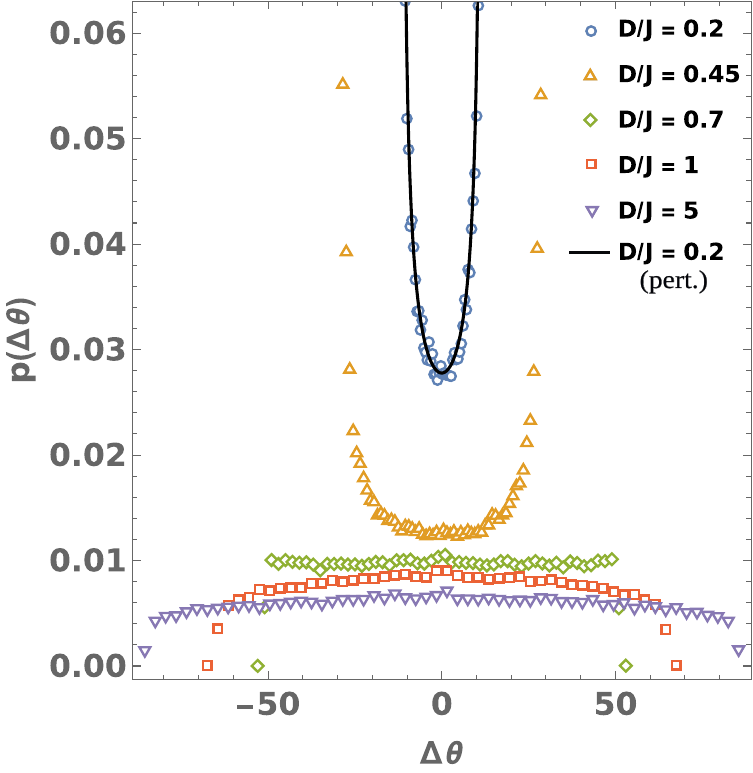}
		\label{fig:hist_d}}
	\subfloat[]{
		\includegraphics[scale=0.53]{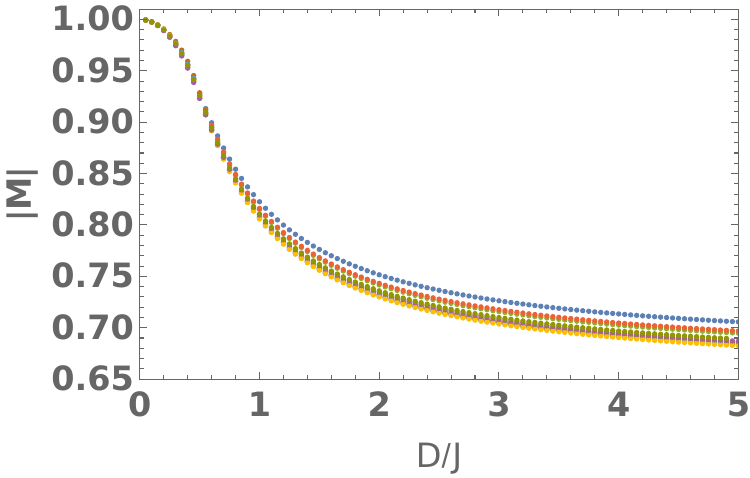}
		\label{fig:mag_vs_D/J}
	}
	\caption{The RCXY model: The sub-figures~(a) and (b) show, respectively, the directions of $\{\boldsymbol{n}_i\}$ and  $\{\boldsymbol{S}_i\}$ for $D/J=0.2$. (c): $\{\boldsymbol{S}_i\}$ for $D/J=5$. All the three sub-figures are for a typical sample with $N=100$. The protruding red dashed (blue straight) arrow shows the direction of $\boldsymbol{n}_0 = \sum_{i=1}^{N} \boldsymbol{n}_i$  ($\boldsymbol{M}$). (d):~Cone angle $\phi$ vs $D/J$ for $10$ different samples with $N=400$. Dashed line depicts $\phi(D/J)$ obtained from the perturbation theory. (e): Probability distribution of deviation of spins $\Delta \theta$ from the direction of magnetization $\theta_0$. The black line shows the distribution~(\ref{probDistnD}) at $D/J =0.2$. (f): Modulus of magnetization vs $D/J$ for $10$ different samples with $N=400$. All the angles are in degrees. }
	\label{fig:DSet}
\end{figure*}

\section{RCXY: Cones at zero-temperature}

We turn to a study of the $T=0$ spin arrangements for the RCXY model as a function of the crystal field strength $D$, using perturbation theory followed by numerical continuation in the non-perturbative regime.

 When $D/J <<1$, the arrangement of spins for the RCXY model is identical to that for the RFXY model. 
  A typical field configuration with $N=100$ is shown in Fig.~\ref{subfig:distn_fields}, which is exactly the same as the configuration  in Fig.~\ref{subfig:betaArrows_100_3} for the RFXY model. Figures~\ref{subfig:coneAt_d0p2_100_3} and \ref{subfig:coneAt_d0p66_100_3} show the corresponding spin configurations at $D/J =0.2$ and $5$, respectively. The red dashed (blue continuous) line shows the direction of vector sum of fields (spins). 
 
 There is a striking difference between the two models for higher values of $D/J$. For instance, the distribution
 of spins within the cone changes and has a maximum at the center of the cone when $D/J \sim 1$, for the RCXY model. Further, unlike the RFXY model, there is no phase transition, and the cone is preserved for all values of $D/J$. Moreover, the cone does not align along $\boldsymbol{n}_0$ in the limit $D/J \rightarrow 0$, in contrast to  the RFXY model. 
 
 The calculation proceeds along the lines spelled out in Sec.~\ref{sec:3}.  
To first order, the perturbative solution for $\{\theta_{i}\}$ that minimize the RCXY Hamiltonian~(\ref{H_with_D}) for small values of $D/J$ is
\begin{align}
	\theta_i = \thetaNot + \frac{D}{J} \left[ 
	\sin(2(\alpha_i - \thetaNot)) - \zeta
	\right],
\end{align}
where $\thetaNot$ is given by
\begin{align}\label{thetaNot_D}
	\tan(2 \thetaNot) = \frac{\sum_{i=1}^N \sin (2 \alpha_i)}{\sum_{i=1}^N \cos (2 \alpha_i)}
\end{align}
and 
\begin{align}
	\zeta = \frac{1}{2} \frac{\sum_{i=1}^{N} \sin \left\{ 4(\thetaNot-\alpha_i)\right\}}{\sum_{i=1}^{N} \cos \left\{ 2(\thetaNot-\alpha_i)\right\}}.
\end{align}
Equation~(\ref{thetaNot_D}) yields four solutions for $\thetaNot$, namely $\widehat{\alpha}, \; \widehat{\alpha} + \pi, \; \widehat{\alpha} + \pi/2 $, and $\widehat{\alpha} - \pi/2$, where $\widehat{\alpha}$ is defined as half of the angle that the vector $\boldsymbol{d}=\left(\sum_{i=1}^N \cos (2 \alpha_i) \right) \boldsymbol{\hat{x}} + \left( \sum_{i=1}^N \sin (2 \alpha_i) \right)\boldsymbol{\hat{y}}$ makes with the $x$-axis.
The first two and the last two solutions are degenerate. However, the latter pair has higher energy and is therefore discarded.

 Note that in the case of RCXY, the solutions for extrema  are always two-fold degenerate as the Hamiltonian is invariant under the transformation $\left\{\boldsymbol{S}_i \, |\, i=1, \; 2, ...N \right\} \rightarrow  \left\{-\boldsymbol{S}_i \, | \, i=1,\;2,...N\right\}$. This implies that there are at least two global minima, and our calculations indicate that there are only two for finite values of $D/J$. In the following, we focus on one of the two minima.

The perturbative solution leads to a magnetization 
\begin{align}
		\boldsymbol{M}  & \equiv \frac{1}{N} \sum_{i=1}^{N} \boldsymbol{S}_i \nonumber \\
		&\simeq \boldsymbol{M}^{(0)} - \frac{D \zeta }{J} \left(
	\sin(\widehat{\alpha}) \boldsymbol{\hat{x}} - \cos(\widehat{\alpha})
	\boldsymbol{\hat{y}} \right)
\end{align}
to first order in $D/J$. It is straightforward to see that $\boldsymbol{M}$ orients along $\theta_0 \simeq \widehat{\alpha} + \frac{D}{J} \zeta$, leading to the following probability distribution for the deviation of a spin $\Delta \theta_i$ from $\boldsymbol{M}$.

	\begin{align}\label{probDistnD}
		p(\Delta \theta_i) = \frac{1}{\pi \sqrt{(D/J)^2 - \Delta \theta_i^2}}.
	\end{align}
Thus, as for the RFXY model, spins are confined within a cone, at whose center the distribution is minimum and at whose edges the distribution diverges. The cone angle is given by $\phi = 2D/J$. 

Beyond the perturbative regime, our numerical studies show that the spins remain confined to a cone, but the distribution becomes flat within the cone at $D/J \sim 0.7$. As $D/J$ is increased further, the distribution develops a maximum at the center of the cone.
Asymptotically, as $D/J \rightarrow \infty$, we expect the distribution to become flat again.  Fig.~\ref{fig:hist_d} shows the probability density $p(\Delta \theta)$ at various values of $D/J$.

\begin{figure}
	\includegraphics[scale=0.9]{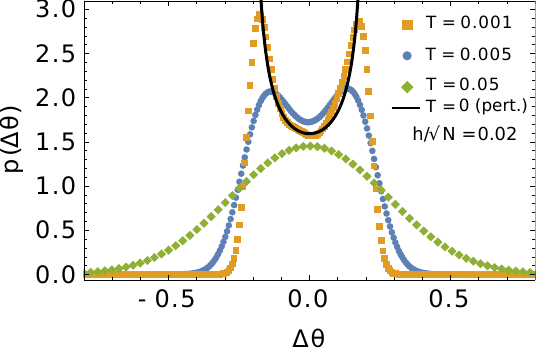}
	\caption{Distribution of spins within the cone for various temperatures, obtained by sampling over $100$ field configurations with $J=1, \; h =0.2$, and $N=100$. The angles are in radians. The distribution for $T=0.05$ is a Gaussian with $0$ mean and standard deviation $0.27$. The black line shows the $T=0, h/J =0.2$ distribution of spins~(\ref{probDistn}), obtained from the perturbation theory. Note that the double-peaked distribution goes to the single-peaked Gaussian distribution when~$T> T_{dis}$.}
	\label{fig:prob_mc}
\end{figure}

The cone angle $\phi$ widens with $D/J$ as shown in Fig.~\ref{fig:phic_vs_d}, initially linearly but the widening slows down when $D/J \sim 1$. As $D/J \rightarrow \infty$, the cone angle $\phi \rightarrow 180^{\circ}$ when $N$ is infinite. This can be conceptualized as follows. 
 In the limit $N \rightarrow \infty$ the fields will be distributed uniformly everywhere over the circle, and  
 as $D/J \rightarrow \infty$ the second term in the Hamiltonian~(\ref{H_with_D}) will dominate forcing each spin $\boldsymbol{S}_i$ to lie on the easy axes pointing either along the direction of the respective field $\boldsymbol{n}_i$ or opposite to it. Thus we have an Ising degree of freedom at each site. However, the $J$ term would prefer the spins to be parallel to each other resulting in the spins spanning a hemisphere. We note here that the hemispherical distribution of spins in the $D/J\rightarrow \infty$ limit has been pointed out in~\cite{Callen_1977}.     

Figure~\ref{fig:mag_vs_D/J} shows the behavior of magnetization as a function of $D/J$. The magnetization decreases smoothly and there is no phase transition. As $D/J \rightarrow \infty$ magnetization $\boldsymbol{M}   \rightarrow 2/\pi$ in the thermodynamic limit~\cite{Barma_2022}.

\begin{figure*}
	\subfloat[]{
		\includegraphics[scale=0.3]{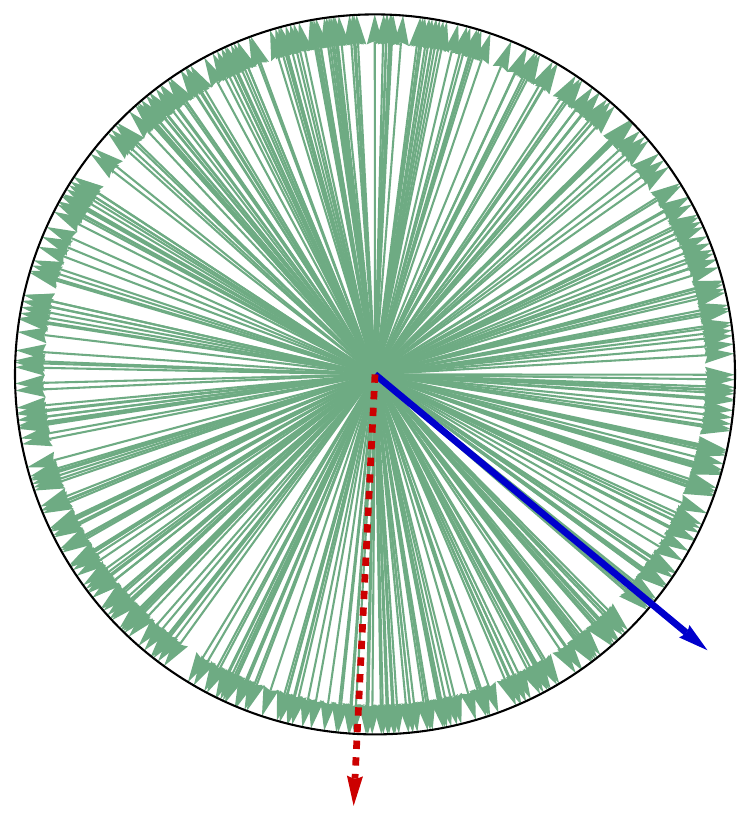}
		\label{subfig:spin_t0}
	}
	\subfloat[]{
		\includegraphics[scale=0.3]{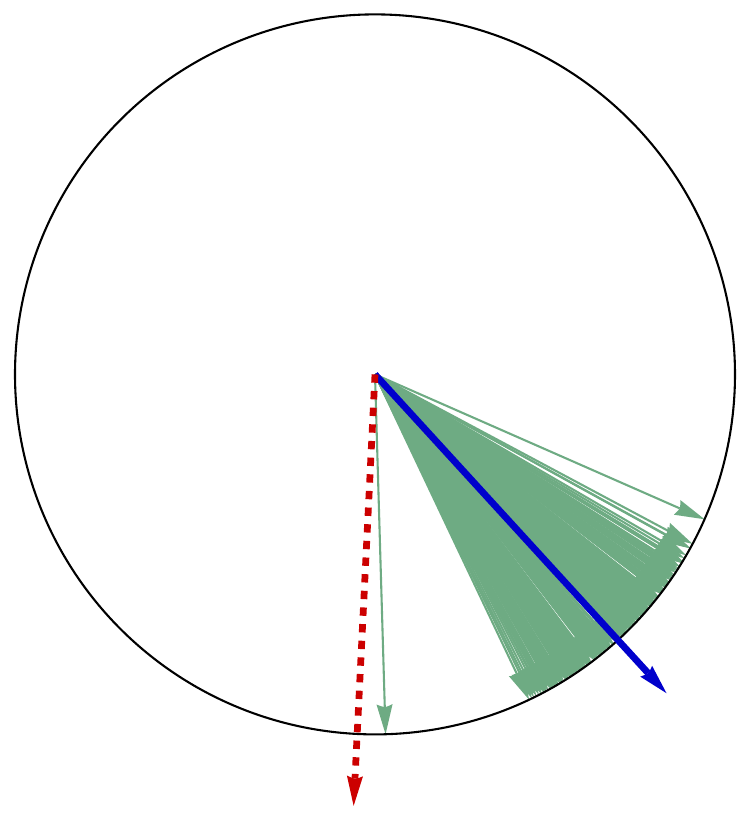}
		\label{subfig:spin_tauf}
	}
	\subfloat[]{
		\includegraphics[scale=0.3]{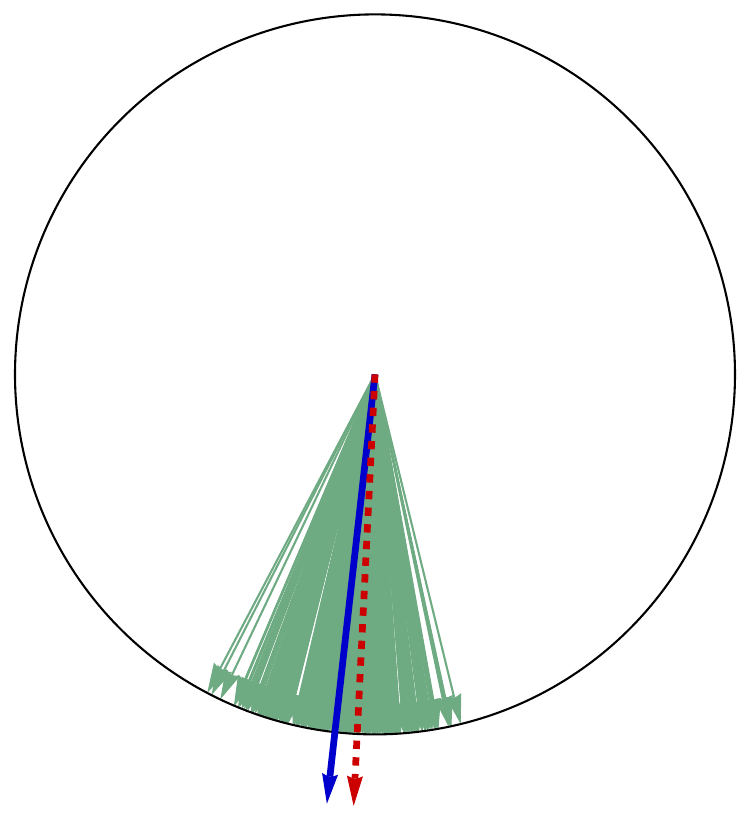}
		\label{subfig:spin_tauo}
	}
	\\
	\subfloat[]{
		\includegraphics[scale=0.8]{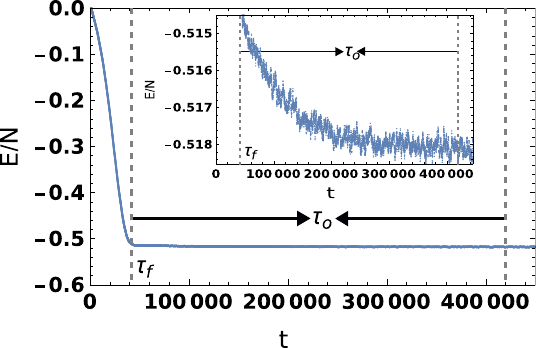}
		\label{fig:evst_mc}}
	\subfloat[]{
		\includegraphics[scale=0.43]{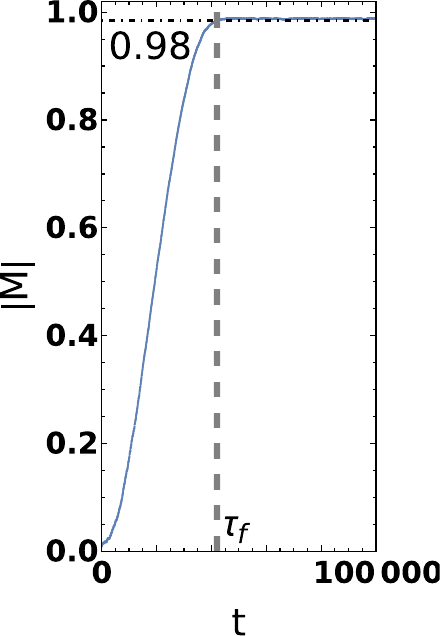}
		\label{subfig:mag_vs_time}
	}
	\subfloat[]{
		\includegraphics[scale=0.8]{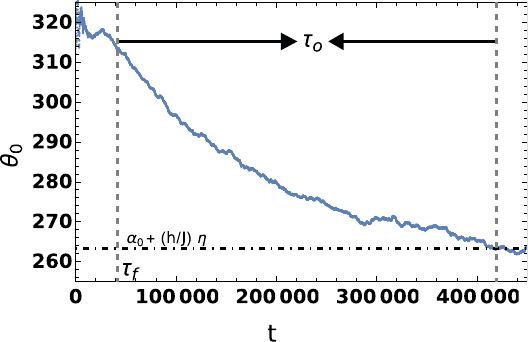}
		\label{t0vst_mc}
	}
\caption{Time evolution (RFXY): Sub-figures (a), (b), and (c) show the spin configurations at $t=0$, $t=\tau_f=4.5 \times 10^{4}$, and $t=\tau_f + \tau_o = 4.35 \times 10^5$, respectively.
		The protruding red dashed and blue straight arrows show the directions of $\boldsymbol{n}_0$ and $\boldsymbol{M}$, respectively.
		(d): Energy per spin versus time. The inset shows variation of $E/N$ from time $\tau_f$ to $\tau_f + \tau_o$.
		(e):  Magnitude of magnetization $\boldsymbol{M}$ versus time.
		(f): Orientation of magnetization $\theta_0$ versus time. The dot-dashed line indicates the $T=0$ orientation of $\boldsymbol{M}$ according to perturbation theory~(\ref{thetaSubNot}). All plots refer to a single typical field configuration and initial spin configuration with $N=400$, $J=1$, $h =0.2$, and $T = 0.005$. }
	\label{fig:MC_set}
\end{figure*}

\section{Cones and dynamics at $T>0$}

Here, we use Monte-Carlo simulations to address the question of robustness of the cone and the spin distribution at equilibrium for the RFXY model at $T>0$. The RCXY model shows similar equilibrium features, and is therefore not discussed separately. We show that the spin distribution  remains similar to that at  $T=0$  and has a two-peaked structure below a characteristic temperature determined by the field. 

We also perform Monte-Carlo simulations below this temperature to study the dynamics of the approach to the ordered spin state starting from a typical random initial configuration of spins. We restrict ourselves to the RFXY model in this case.

\subsection{Equilibrium cones}

To first order in $h/J$, the energy per spin in the ground state is given by
\begin{align}\label{pert_eh}
	E/N = -J/2 -\frac{h}{N}  \sum_{i=1}^{N} \cos(\alpha_0 - \alpha_i) = -J/2 - \frac{h}{N} |\boldsymbol{n}_0|,
\end{align}
which is obtained by using Eqs.~(\ref{H_with_h}) and (\ref{soln_perth}).
  The contribution coming from the randomly disordered fields is $E_{dis} = h|\boldsymbol{n}_0|/N$. Since $\boldsymbol{n}_i$'s are independent random vectors distributed uniformly over a unit circle, we have 
\begin{equation}
	|\boldsymbol{n}_0| \equiv \left|\sum_{i=1}^{N} \boldsymbol{n}_i \right| \sim \sqrt{N},
\end{equation}
implying, $E_{dis} \sim h/\sqrt{N}$. Thus disorder brings into play a characteristic temperature $T_{dis} \sim h/\sqrt{N}$. We performed Monte-Carlo simulations at $T<T_{dis}$ and $T>T_{dis}$ according to the following procedure. 

 First, a random field configuration is realized by choosing $N$ field-angles $\{\alpha_i\}$ independently and uniformly from the interval $\left(\left. 0, 2 \pi \right.\right]$. An initial spin configuration is chosen by assigning  a set of $N$ angles $\{\theta_i\}$ in the same fashion.  The angles $\{\theta_i\}$ are now evolved in steps using the Metropolis algorithm. Each step consists of $N$ micro-steps, at each of which a particular $\theta_i$ is chosen at random and varied by an angle $\Delta \theta_i$ that is chosen randomly and uniformly from the interval $(-\pi/500, \pi/500)$. This changes the spin angles, say, from $\{\theta_i\}$ to $\{\theta'_{i}\}$. At the end of a step, if the new spin configuration lowers the energy~(\ref{H_with_h}), it is accepted. Otherwise it is accepted with a probability $\exp(-\Delta H_{RF}/T)$, where $\Delta H_{RF} = H_{RF}\boldsymbol{(}\{\theta'_{i}\}\boldsymbol{) }-  H_{RF}\boldsymbol{(}\{\theta_{i}\}\boldsymbol{) }$. Each such step corresponds to one unit time.     

 From the simulations, we find that the equilibrium distributions of spins within the cone have different characteristics for $T<T_{dis}$ and $T>T_{dis}$, as shown in Fig.~\ref{fig:prob_mc}. 
  For $T < T_{dis}$, the distribution is similar to that at $T=0$,  with  maxima towards the edges of the cone and minimum at the center, whereas  for $T>T_{dis}$, the distribution is a Gaussian peaked at the center of the cone.  The equilibrium distributions of spins exhibit the same features for the RCXY model also, with the characteristic temperature $T_{dis} \sim D/\sqrt{N}$.

\subsection{Low temperature dynamics}

We now study the dynamics of spins in the ordered state using analytical and Monte-Carlo methods, restricting ourselves to $T<T_{dis}$.  

We find that the evolution of spins from a typical random initial configuration to the equilibrium state is characterized by two time scales. In the initial, shorter time scale $\tau_f$, the randomly oriented spins come together and form a cone, while on the second longer time scale $\tau_o$, the cone rotates and orients along the equilibrium direction. Figure~\ref{fig:MC_set} shows illustrations of evolution of the spin configuration, and plots depicting the evolution of the magnitude and the orientation of magnetization and the evolution of energy  for a typical field configuration and initial spin configuration. 

\begin{figure}
		\includegraphics[scale=0.65]{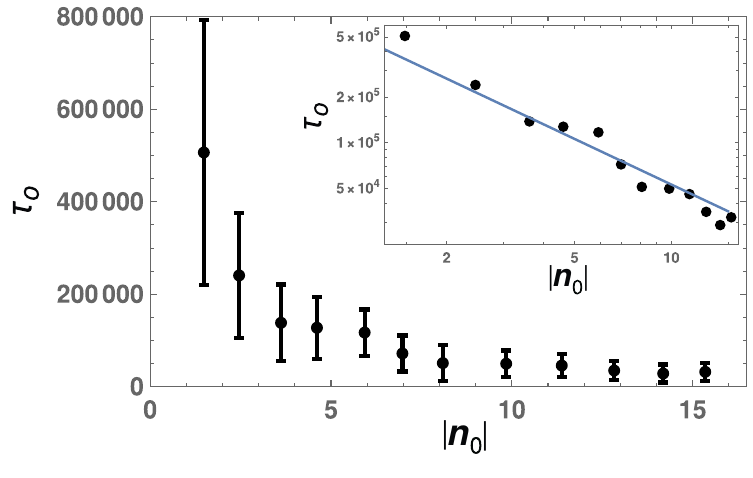}
		\caption{The cone orientation time $\tau_o$ for different field configurations, each with a different value of $|\boldsymbol{n}_0|$. For each field configuration, $\tau_o$ is calculated by averaging over $100$ different initial spin configurations. The height of the error bar shows the standard deviation of $\tau_0$ with initial spin configurations. The inset shows $\tau_0$ versus $|\boldsymbol{n}_0|$ in $\log$ scale. The blue line therein has slope~$-1$. Note that $N=100$ here.}
		\label{fig:tauo}
\end{figure}

We find that the cone formation time $\tau_f$ depends primarily on $J$, while the cone-orientation time $\tau_0$ is set by $h$. Further, $\tau_0$ depends on $	|\boldsymbol{n}_0| \equiv \left|\sum_{i=1}^{N} \boldsymbol{n}_i \right| $. The larger the value of $|\boldsymbol{n}_0|$, the shorter is the orientation time. This is evident from Fig.~\ref{fig:tauo}, which shows the cone orientation time $\tau_o$ for different field configurations, each with a different value of $|\boldsymbol{n}_0|$. Note that in practice, we calculate $\tau_f$ as the time at which the value of $|\boldsymbol{M}|$ first reaches $0.98$ and $\tau_f +\tau_o$ as the time at which $\boldsymbol{M}$  orients at an angle $\alpha_0 + (h/J) \eta$, which is the $T=0$ orientation of $\boldsymbol{M}$ according to the perturbation theory~(\ref{thetaSubNot}).  

We write a phenomenological  dynamical equation to describe the evolution of the orientation of $\boldsymbol{M}$ once the cone has formed:  
\begin{equation}\label{dynamics_h}
	\frac{\partial \theta_0}{\partial t} = -\gamma \frac{\partial H_{RF} }{ \partial \theta_0}, 
\end{equation}
where the effects of thermal fluctuations are neglected.
 Since we are interested primarily in the orientation of the cone, we neglect the spread of spins within the cone and write $\boldsymbol{S}_i = \boldsymbol{S}$ for all $i$. This simplification is justified for small enough $h/J$ when the spins point more or less along the same direction once the cone has formed. Thus, Eq~(\ref{H_with_h}) becomes
\begin{align}
	H_{RF} &= - \frac{J}{2 N} \left(N \boldsymbol{S} \right)^2 - h \sum_{i=1}^N \boldsymbol{S} \cdot \boldsymbol{n}_i 
	\\ 
	&= -\frac{JN}{2} - h \sum_{i=1}^N \cos(\theta_0 - \alpha_i),
	\label{simham}
\end{align} 
where $\theta_0$ denotes the orientation of $\boldsymbol{S}$. 
   Defining $\Delta \theta_0 = \theta_0 - \alpha_0$, the sum in the second term in Eq.~(\ref{simham}) can be rewritten as 
\begin{align}
	\sum_{i=1}^N \cos(\theta_0 - \alpha_i) =& \sum_{i=1}^N \cos(\alpha_0 - \alpha_i + \Delta \theta_0) \nonumber \\
	=& \cos(\Delta \theta_0) |\boldsymbol{n}_0|,
	\label{simcos}  
\end{align}
where we used the results $\sum_{i=1}^{N} \cos(\alpha_0 -\alpha_i) = |\boldsymbol{n}_0|$ and $\sum_{i=1}^{N} \sin(\alpha_0 -\alpha_i) =0$. Solving for $\theta_0$ using Eqs.~(\ref{dynamics_h}), (\ref{simham}), and (\ref{simcos}), we obtain

\begin{equation}
	\tan\left( \frac{\theta_0(t) - \alpha_0}{2} \right) = \tan \left( \frac{\theta_0(0) - \alpha_0}{2} \right) \exp\left( - \gamma h |\boldsymbol{n}_0| t \right).
\end{equation}
From the above we see that the relaxation time~$\tau_o \propto {1/|\boldsymbol{n}_0|}$ which is corroborated by the numerical results (see Fig.~\ref{fig:tauo}). 

\section{Conclusion}

The ordered states in the RFXY and the RCXY models show  interesting arrangements of spins, which break rotational symmetry for any finite $N$. In the ground state, the spins are confined within a two-dimensional cone, whose angle and orientation are determined by the ratio of disorder to coupling strength. The distribution of spins within the cone is sensitive to temperature, and shows very different features for $T<T_{dis}$ and $T>T_{dis}$. The dynamics of spins relaxing to equilibrium is characterized by two time scales, namely, the cone formation time and the cone orientation time. 

Note that rotational symmetry is restored only if $N$ is strictly infinite: the ground state energy per spin $E/N$ becomes insensitive to the orientation of $\boldsymbol{M}$ in that limit. Further,  our preliminary analysis of the dynamics suggests that $\gamma \sim N^{-\alpha}$, where $\alpha>0$, implying that the cone orientation time $\tau_o$ diverges as $N\rightarrow \infty$. The cone formation time is also sensitive to $N$. It would be interesting to track the $N$-dependence of these times.

Another physically relevant question is: what is the effect of a uniform external field on the cones and arrangement of spins, as well as on the phase diagram? A preliminary study, based on the extension of our numerical continuation techniques to this problem, suggests that at $T=0$, the external field reorients the cone, with a possible first order phase transition --- from an ordered state to a different ordered state.     

We  expect many of the qualitative features of the RFXY and RCXY models, like the existence of a cone, to remain valid for the case of Heisenberg spins $S_i$ that can lie anywhere on a unit sphere.  

\section{Acknowledgment}
We thank Arghya Das, Smarajit Karmakar, and Sumedha  for fruitful discussions. M.B. acknowledges the support of the Indian National Science
Academy(INSA). 
We acknowledge the support of the Department of Atomic
Energy, Government of India, under Project Identification No. RTI 4007.

\appendix

\section{Numerical continuation}\label{sec:app1}
Here, we explicitly lay down the procedure for solving the nonlinear Eqs.~(\ref{H_with_h_extrema}) numerically.
 
The left hand side of Eq.~(\ref{H_with_h_extrema}) $G_i \equiv \partial H_{RF}/\partial \theta_i$, for $i=1, \;2, ... N$, 
constitute a set of continuously differentiable functions of the variables $\boldsymbol{\theta} = \{\theta_1, \; \theta_2, \; ... \theta_N \}$ and the parameter $\lambda \equiv h/J$, for a fixed $N$ and  field angles $\{\alpha_i \}$. It is useful to write down Eq.~(\ref{H_with_h_extrema})  in terms of $G_i$ here: 
\begin{equation}\label{main}
	G_i=0.
\end{equation}
If for a set of values $(\thetab,\lambda)$, $G_i ( \thetab, \lambda) = 0$ for every $i$, then that set is called a solution-point.

If there exists a solution-point $(\thetab_0, \lambda_0)$ for which 
 \begin{equation}\label{reg_cond}
\det\left[
\begin{matrix}
	\frac{\partial G_1}{\partial \theta_1} & ... & \frac{\partial G_1}{\partial \theta_N}\\
	\vdots &   &   \\
	\frac{\partial G_N}{\partial \theta_1} & ... &\frac{\partial G_N}{\partial \theta_N} \\
\end{matrix}
\right]_0 \ne 0,
\end{equation}
where the subscript $0$ indicates that the derivatives are to be taken at $(\boldsymbol{\theta}, \lambda) =(\boldsymbol{\theta}_0, \lambda_0)$, then there also exists a unique set of solution-points $(\thetab(\lambda), \lambda)$ for values of $\lambda$ in the neighborhood of $\lambda_0$, by implicit function theorem. Such a solution point is referred to as regular.

If we know a regular solution-point $(\thetab_0, \lambda_0)$, we can find a nearby solution-point $(\thetab_1, \lambda_1)$ approximately by incrementing the parameter $\lambda$ by a small number $\Delta \lambda$ to obtain $\lambda_1 = \lambda_0 + \Delta \lambda$ and then finding $\thetab_{1} = \thetab_0 + \Delta \thetab$ that solves the linearized version of Eq.~(\ref{main}), which is 
\begin{equation}
	\sum_{j=1}^{N} \frac{\partial G_i(\thetab_0, \lambda_0) }{\partial \theta_j} \Delta \theta_j + \frac{\partial G_i(\thetab_0, \lambda_0) }{\partial \lambda} \Delta \lambda = 0.
\end{equation}
The above equation can be solved easily to obtain $\Delta \thetab$, and hence $\thetab_{1}$. The accuracy of the approximate solution $\thetab_1$ can be improved by using any of the familiar iterative schemes, such as the Newton-Raphson or the Secant method. In this paper we have used the inbuilt FindRoot function in Mathematica. 

If the new solution point $(\thetab_1, \lambda_1)$ is also regular, we can repeat the above scheme to find the next solution point $(\thetab_2, \lambda_2)$, and so forth. Alternate methods have to be used if a solution-point is not regular. For solving Eq.~(\ref{H_with_h_extrema}), we start from the perturbative solution~(\ref{soln_perth}) and not from the solution $\thetaNot$ at $h/J =0$ because the latter is not a regular solution-point. However, the subsequent solution points that correspond to the minima of the Hamiltonian~(\ref{H_with_h}) are regular, and therefore no alternate schemes had to be used. Further, the absence of non-regular solution-points implies that there are no bifurcations or folds at any of these points.

\bibliography{references}
\end{document}